\documentclass[10pt,twocolumn]{article} 
\usepackage{latex8}
\usepackage[latin1]{inputenc} \usepackage[T1]{fontenc}
\usepackage{textcomp} \usepackage[british,german]{babel}
\usepackage{amsmath,xspace}
\newif\ifpdf \pdffalse 
\usepackage[dvips]{graphicx}
\newcommand{\ie}{, {i.e.},\xspace}
\newcommand{\eg}{, {e.g.},\xspace}

\begin{document}
\selectlanguage{british}
\title{Employing Trusted Computing for the forward pricing of pseudonyms in reputation systems}

\author{Nicolai Kuntze, Dominique Mähler, and Andreas U. Schmidt\\
Fraunhofer--Institute for Secure Information Technology SIT\\ 
Rheinstraße 75, 64295 Darmstadt, Germany\\
Andreas.U.Schmidt@sit.fraunhofer.de}
\maketitle
\thispagestyle{empty}

\begin{abstract}
Reputation and recommendation systems are fundamental for the formation
of community market places.
Yet, they are easy targets for attacks which disturb a market's equilibrium
and are often based on cheap pseudonyms used to submit ratings.
We present a method to price ratings using trusted computing, based
on pseudonymous tickets.
\end{abstract}
\Section{Introduction}
Market places for virtual goods are increasingly occupied by self-organising
communities. These market places exhibit the characteristics of the
so-called long tail economy. That is, the classical asymmetry between 
suppliers and consumers is lifted. Buyers and sellers are often even in numbers
and may change their roles dynamically. Virtual, or physical, 
goods are offered in large numbers and diversity and with potentially small
demand for each single one.
Matchmaking and orientation of buyers is difficult in a long tail economy,
long term relationships are hard to build, and trust between trade partners
must be established somehow~\cite{Bakos1998}.

A common approach is to let market players themselves provide the necessary 
guidance. This is mostly embodied in  reputation systems by which buyers and 
sellers rate each other and the goods sold, or recommendation systems\ie  
programs which attempt to predict items that a user may be interested in, 
given some information about the user's profile.
Reputation systems, according to Paul Resnick \textit{et al.}~\cite{Resnick2000}
``seek to establish the shadow of the future [the expectation of reciprocity
or retaliation in future interactions, cf.~\cite{Axelrod1984}] 
to each transaction by creating an expectation that other people will look back on it''. 
The goal is to establish a homogeneous market for honest participants.
That community ratings (of goods) do in fact strongly influence buyer 
behaviour is shown empirically in~\cite{SDW06}. 

Existing reputation systems are fragile, in that they can easily 
be distorted or abused even within the frame of laws governing them. 
`Attacks' of this kind, though not proper attacks in the sense of 
information security, threaten the integrity --- with respect to its purpose ---
of the informational content stored in the system.
Dellarocas~\cite{Dellarocas00} clssifies unfair bahaviour 
into the categories
1.~\textit{Ballot stuffing}: A seller colludes with a group of buyers in order to 
be given unfairly high ratings.
2.~\textit{Bad-mouthing}: Sellers and buyers collude to rate other sellers
unfairly low to drive them out of the market.
3.~\textit{Negative discrimination}: Sellers provide good services only to
a small, restricted group of buyers.
4.~\textit{Positive discrimination}: Sellers provide exceptionally good service to
some buyers to improve their ratings.
A situation of controlled anonymity in which the market place knows the identity of
paricipants and keeps track of all transactions and ratings, but conceals the identity
of buyers and sellers, is identified as essential to avoid unfair behaviour.
For instance, anonymity is an effective protection against bad-mouthing, but 
cannot work for ballot stuffing as sellers can give hidden indications of their
identities to colluders.

On the other hand, the best known individual attack on reputation systems
uses Sybils to obtain a disproportionately large influence~\cite{Douceur2002}.
Friedman and Resnick~\cite{Friedman2001} point to the general problem of 
`cheapness' of pseudonyms in marketplaces and reputation systems, since with name 
changes dishonest players easily shed negative reputation, as corroborated 
theoretically in~\cite{Dellarocas2004}. The paper~\cite{Bhatta2005} gives an explicit threshold 
for the transaction costs for reputations needed to avoid ballot stuffing. 
However, an indiscriminate pricing
of identities for the submission of ratings poses an undesired entry deterrent.
It seems therefore plausible that reputation systems should be based on pseudonyms
which allow for a flexible forward pricing.

While related work addresses particular vulnerabilities~\cite{Friedman05}, or proposes general
frameworks to ensure accountability in reputation systems, while maintaining
anonymity~\cite{Buttyan1999,Zieglera2006}, we here propose a simple mechanism to
introduce arbitrary costs for pseudonyms.
The principal approach is to use identities provided by trusted computing (TC)
mechanisms for this purpose.
The next section introduces the essentials of trusted computing necessary to
develop our concept.
In Section~\ref{sec:Concept}, the central concepts for 
architectures and protocols for acquisition and usage of TC-based identities
in reputation systems is described. 
Section~\ref{sec:app} shows how this is used in an integrated scenario to
price reputations.
%
\Section{Trusted Computing essentials}\label{sec:TCess}
TC aims at establishing trust in computing platforms.
The Trusted Computing Group (TCG) has defined a family of standards~\cite{TPM06} to
this end. A so called Trusted Platform Module (TPM) is the
trust anchor offering the ability to securely create, store, and use asymmetric keys.
Moreover, a TPM in cooperation with an appropriate operating environment can
issue to a third party assertions about the trustworthiness of the computing
platform\eg signal that the platform is in a secure state. Such a TPM
enhanced system is called trusted platform.

The TPM has so-called shielded capabilities protecting internal
data structures by controlling their use. For the present application we
use two particular features of a trusted platform. First, key creation and management, 
and second the ability to create a trust measurement which can be used to assert 
a certain state toward a third party. 
The TPM is equipped with a physical random number generator, and a key generation
component which creates RSA key pairs. The key generator is designed as a
protected capability and the created private keys are kept in a shielded
capability (a protected storage space \textit{inside} the TPM). 

A crucial concept are the so called attestation identity keys (AIK) which
are used to sign the trust measurements or to certify keys. 
AIKs can be used, according to TCG standards, to attest the 
origin and authenticity of a trust measurement, but it can also
authenticate other keys generated by the TPM. The latter functionality 
is central for our intended application.
Before an AIK can testify the authenticity of any data, a Privacy Certification
Authority (PCA) has to issue a credential acknowledging that this AIK belongs
to a certain TPM which is deployed in a trusted platform. 
The protocol for issuing this credential consists in three basic steps. 
First, the TPM generates an RSA key pair by performing the \verb+TPM_MakeIdentity+
command. The resulting public key together with certain credentials  identifying
the platform is then transferred to the PCA. Second, the PCA 
verifies the correctness of the produced credentials and the AIK signature. If
they are valid the PCA creates the AIK credential which contains an identity
label, the AIK public key, and information about the TPM and the platform.
A special structure containing the AIK credential is created which is
used in step three to activate the AIK by executing the \verb+TPM_ActivateIdentity+
command. So far, the TCG-specified protocol is not completely secure, since
between steps two and three, some kind of handshake between PCA and platform is
missing. The protocol could be enhanced by a challenge/response part to
verify the link between the credentials offered in step one and used
in step two, and the issuing TPM. 
\Section{Pseudonymous rating with AIKs} \label{sec:Concept} 
The basic idea is to establish a pseudonymous rating system using the
identities embodied in the PCA-certified AIKs.
We first describe how AIKs can be turned into tickets that can be used
in a reputation system and then develop the processes for their 
acquisition and redemption.
\SubSection{AIKs as rating tickets}\label{sec:AIKusage}
For security considerations the TPM restricts the usage of AIKs. 
It is not possible to use AIKs as signing keys for arbitrary data and in particular 
to establish tickets in that way.
It is therefore necessary to employ an indirection using a TPM generated signing key
and certify this key by signing it with an AIK --- \textit{viz} \textit{certify} it
in the parlance of the TCG. 
Creation of a key is done by executing the
\verb+TPM_CMK_CreateKey+ command, which returns an asymmetric key pair where the
private portion is encrypted by the TPM for use within the TPM only. 
The resulting
key pair is loaded into the TPM by \verb+TPM_LoadKey+
and thereafter certified by \verb+TPM_CertifyKey+. 
By certifying a specific key the TPM makes the statement that
``this key is held in a TPM-shielded location, and it will never be
revealed''. For this statement to have veracity, a challenger or verifier  must
trust the policies
used by the entity that issued the identity and the maintenance policy of
the TPM manufacturer. 

This indirection creates to each AIK a certified key (by the namely AIK)
that can be used for signing data, in particular the payload of a rating
to be submitted to a reputation system.
We call this key pair the \textit{certified signing key} (CSK).
CSK, AIK, together with a certificate by the PCA (see below)
attesting the validity of that AIK, are the ingredients that realise a ticket
for submission of a single rating.
\SubSection{Ticket acquisition and rating process}\label{sec:process}
Rating tickets are acquired by a \textit{trusted agent} (TA)\ie the user of a
reputation system operating with his trusted platform, from the PCA.
They are then redeemed at the \textit{reputation system} (RS). In both, a 
\textit{charging provider} (CP) may occur as a third party. We now describe how 
these operations proceed.
Note that we do not distinguish between public and private key portions of a
certificate establishing a credential. As a notation, 
the credential of some certified entity 
$\text{Cert(\textit{entity}, \textit{certificate})}$ means the union of
the public key $\text{Pub(\textit{certificate})}$ and the entity signed
with the certificate's private key, $\textit{entity}_{\text{Priv}(\textit{certificate})}$.
Verifying a credential means to check this digital signature.

The credentials issued by the PCA for a AIKs are \textit{group credentials}\ie
they do not identify a single AIK \textit{viz} ticket but rather its price or
value group $g$ chosen from a predetermined set indexed by the natural numbers $g\in\{1,\ldots,G\}$.
The group  replaces an individual identity of a platform and many TAs will get the
same group certificate. Only the PCA can potentially resolve the individual identity
of a platform.
These groups are used to implement price and value discrimination of ratings.
Note that the PCA is free in the choice of methods to implement group certificates.
This could be done by simply using the same key pair for the group or by the existing,
sophisticated group signature schemes~\cite{Chaum91}.

\begin{figure}
\centering 
\ifpdf\resizebox{0.45\textwidth}{!}{\includegraphics{Ticket_Acquisition.pdf}} \else
  \resizebox{0.45\textwidth}{!}{\includegraphics{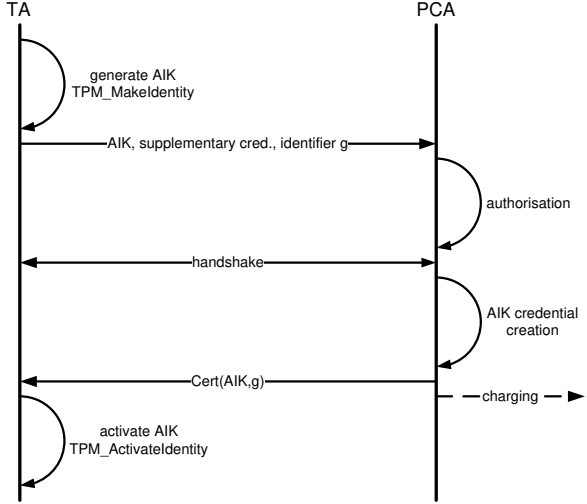}} \fi
  \caption{Ticket acquisition process.}
  \label{fig:Ticket_Acquisition}
\end{figure}

If a TA wants to acquire a rating ticket from group $g$, he first generates an 
AIK using the \verb+TPM_MakeIdentity+ command.
Next, TA requests from the PCA a credential for this AIK, belonging to group $g$,
by sending AIK, group identifier and supplementary data as required by TCG protocols,
to the PCA.
The PCA now knows the identity of the TA.
This can be used to perform a charging for the ticket, 
either by contacting CP or by the PCA itself
(how charging actually works is not in the scope of this paper).
It is important that at this stage an authorisation decision on the ticket generation
can be made by the PCA, for instance to blacklist misbehaving participants.
If the authorisation succeeds (and not earlier to save bandwidth and resources), 
the PCA performs a handshake operation with the TA to ensure that the AIK has actually
been generated by the particular TPM in question.
Upon success, the PCA generates the credential
$\text{Cert}(\text{AIK},g)$ certifying that the AIK belongs to group $g$.
The credential is transferred back to TA, where finally the 
\verb+TPM_ActivateIdentity+ command is executed to enable subsequent usage of this AIK.
The process is shown in Figure~\ref{fig:Ticket_Acquisition}.

Redeeming a rating ticket and submitting a rating $r$ is now very simple.
TA has first to generate a CSK\ie a public/private key pair and
the credential $\text{Cert}(\text{CSK},\text{AIK})$ for it according
to the process described in Section~\ref{sec:AIKusage}.
He then signs $r$ with CSK to obtain $\text{Cert}(\text{r},\text{CSK})$.
The rating and the credential chain
$\text{Cert}(\text{r},\text{CSK})$, $\text{Cert}(\text{CSK},\text{AIK})$, 
$\text{Cert}(\text{AIK},g)$
is then transferred from TA to RS.
RS verifies this chain and makes an authorisation decision, for instance
to implement a protection against multiple spending.
Finally, RS acknowledges receipt of $r$ and optionally
initiates another charging operation (ex post charging) 
via PCA.
\begin{figure}
\centering 
\ifpdf\resizebox{0.45\textwidth}{!}{\includegraphics{Ticket_Redemption.pdf}} \else
  \resizebox{0.45\textwidth}{!}{\includegraphics{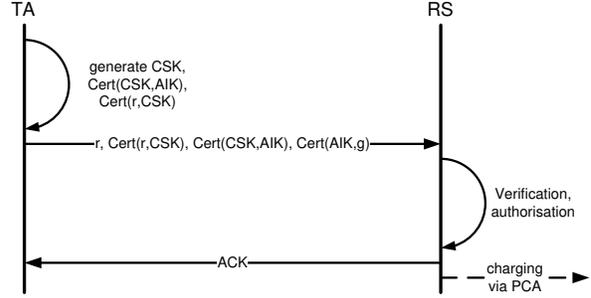}} \fi
  \caption{Ticket redemption process.}
  \label{fig:Ticket_Redemption}
\end{figure}
\SubSection{Security and anonymity}\label{sec:sec}
The presented method for the management of ratings in a reputation system
provides for perfect pseudonymity of the participants toward the system.
In fact, only PCA is able to de-anonymise users.
This implements precisely the controlled anonymity desired for reputation systems.
Note that for the namely reason only PCA can initiate a charging, since only
he knows (or is able to know) the identity of a TA and can link
it to the identity of the corresponding participant.

To keep this pseudonymity strength, it is essential that our concept relies
only on genuine TPM functionality, and in particular avoids the usage
of trusted software.
If there was a trusted software managing rating tickets in some way at the side of TA, 
then this software, and the state of the platform would have to be attested both
in ticket acquisition and redemption.
To this end the TC protocols for remote attestation transfers trust  measurements and
measurement logs to the corresponding verifier (PCA or RS in our case).
These data can however --- and this is a principal problem with remote attestation ---
be used to individualise the trusted platform, if, as in the PC domain,
the number of system states and different measurement logs created at boot time,
is very large in relation to the number of users of a TC-based service.
Besides, avoiding remote attestation saves bandwidth and resources consumption. 
Since no trust can be laid in the TA for rating ticket management,
some kind of double, or multiple, spending protection or 
usage authorisation is needed at RS upon ticket redemption.
\Section{Application scenario}\label{sec:app}
\begin{figure}
\centering 
\ifpdf\resizebox{0.47\textwidth}{!}{\includegraphics{Rating_Scenario.pdf}} \else
  \resizebox{0.47\textwidth}{!}{\includegraphics{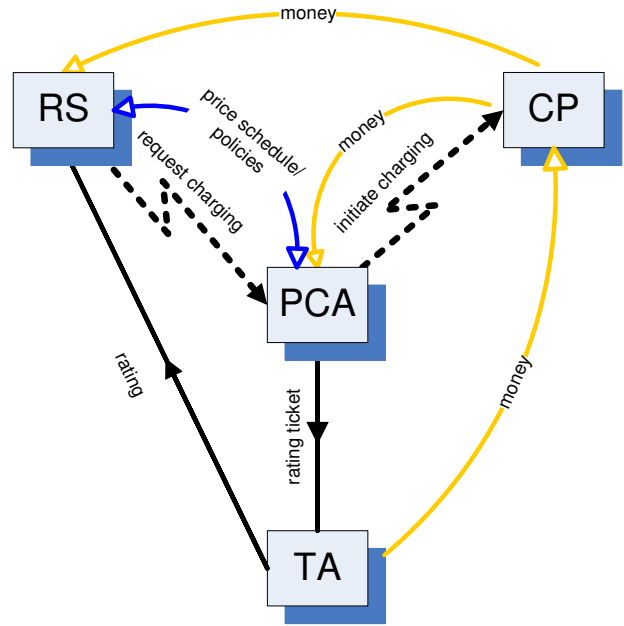}} \fi
  \caption{A reputation system with PCA.}
  \label{fig:Rating_Scenario}
\end{figure}
The embedding of the described ticket acquisition and redemption into a (commercial) 
reputation system or another business context offers many variants.
A very basic scenario is shown in Figure~\ref{fig:Rating_Scenario}.

Here, a trusted agent (user) would like to express a rating about another user 
(for example, a buyer about a seller, a seller about a buyer), and would buy a  rating 
ticket from the PCA.
The ticket belongs to a certain group which can represent statements such as ``for usage with 
RS XYZ'' (for example, eBay), and a certain rating level, e.g., an impact factor
used by the RS to calculate weighted overall ratings.
The user then formulates the rating and send it to the RS.
The TA pays for the ticket at the CP at the time of redemption of the ticket and the
CP distributes revenue shares between himself,
PCA, and RS, according to service level agreements.
The result would be a rating statement about another participant of the RS which is trustworthy, 
because it is non-repudiatable by the signature chain, 
accountable by resolution of the TA's identity through PCA
but protected as a pseudonym by the mechanism presented in this paper.
 
The PCA plays a very central role for the control of identities embodied
in the pseudonymous tickets that it issues. It is in fact similar to the
role of an identity provider in identity management (IDM).
That TC can be used to model IDM was outlined in~\cite{KuntzeSchmidt2006A,KuntzeSchmidt2006C}, 
and is exemplified in the present paper for the first time.
Though the separation of duties between PCA and CP allows in principle even
for anonymity of the person using a TA in the RS, since only upon charging
this person must be identified by credit card account or other means, this
may not be the best option.
One important problem in reputation systems is accountability of users\ie
the possibility to trace back malicious ones and threaten them with personal consequences.
While RS may be able to obtain such personal identities from PCA in such
cases if pertinent contractual relationships are in place, 
data protection policies or regulations may prevent a CP from unveiling personal
identities (if no fraud or monetary damage is suspected).

The second role played by the PCA is for initiation of charging for tickets
and therefore for the intended pricing of ratings.
With respect to the revenues from rating ticket sales, a natural approach
seems to be a sharing between RS and PCA (and CP for its service).
RS and PCA negotiate and implement policies for authorisation within
the ticket acquisition and redemption processes\eg to prevent double spending
or to blacklist misbehaving users.
In collaboration between PCA and RS, practically any pricing scheme for rating
tickets can be realised. 
On the extreme ends of the spectrum are cost-free registration of 
ratings by PCA, ensuring only accountability, and increasing charges with the
number of ratings (or\eg their frequency).
Even reverse charging\ie paying incentives for ratings\eg such of good quality,
is possible.
\Section{Conclusions}
Though envisaged with reputation systems as the main motivation
it is clear that the ticket acquisition and redemption system
above can be extended to a generic pseudonymous ticket system for arbitrary
service access or the acquisition of virtual goods. In fact, what we have 
constructed is essentially a payment system with a trusted 
third party guaranteeing pseudonymity.

The presented scheme also extends to an arbitrary number 
of reputation systems to which the PCA offers rating ticket pricing as
services. A further extension would be to let TA express values of 
tickets by using different (groups of) CSKs. In this way ratings could be prioritised.

It is worthwhile to compare our method with the use case scenario ``Mobile payment''
of TCG's Mobile Phone Working Group Use Case Scenarios~\cite[Section 8]{MPWGUC05}.
There, the focus lies on device-side support of payment operations on a mobile
phone which is turned into a trusted platform. This always involves a trusted
software on the device which is not required in our approach.
On the other hand this is only possible through the introduction of a trusted 
third party, the PCA with its extended duties.
Thus we lack the universality of client-side solutions.
Yet we have shown that a very simple ticket system with strong pseudonymity
can be established resting solely on the most basic TPM functions.
\Section{Acknowledgements}
The authors thank an anonymous referee for his constructive comments.
\providecommand{\noopsort}[1]{} \providecommand{\singleletter}[1]{#1}

\end{document}